\documentclass[sigconf,natbib]{acmart}

\copyrightyear{2018} 
\acmYear{2018} 
\setcopyright{acmlicensed}
\acmConference[SIGIR '18]{The 41st International ACM SIGIR Conference on Research \& Development in Information Retrieval}{July 8--12, 2018}{Ann Arbor, MI, USA}
\acmPrice{15.00}
\acmDOI{10.1145/3209978.3210031}
\acmISBN{978-1-4503-5657-2/18/07}

\usepackage[subtle, wordspacing=normal, tracking=normal, bibnotes, charwidths=normal, leading, indent, lists, paragraphs=normal, mathspacing=normal, bibliography=normal]{savetrees}
\usepackage{tabularx}
\usepackage{booktabs}
\usepackage[skip=0pt]{caption}
\usepackage[skip=0pt]{subcaption}
\usepackage{pgf, tikz}
\usetikzlibrary{arrows,automata,shapes}
\usetikzlibrary{positioning}
 \usepackage[noend]{algpseudocode}
\usepackage{enumitem}
\usepackage{amsmath}
\usepackage{graphicx}
\usepackage{algorithm}

\usepackage{color}

\newcommand{\todo}[1]{\textcolor{red}{[#1]}}
\newcommand{\NV}[1]{\textcolor{purple}{[#1]}}

\title{Weakly-supervised Contextualization of Knowledge Graph Facts}

\author{Nikos Voskarides} 
\authornote{This work was done while Nikos was an intern at Bloomberg.}
\orcid{}
\affiliation{%
\institution{University of Amsterdam}
\city{Amsterdam}
\country{The Netherlands}
}
\email{n.voskarides@uva.nl}

\author{Edgar Meij}
\orcid{}
\affiliation{%
\institution{Bloomberg}
\city{London}
\country{U.K.}
}
\email{edgar.meij@acm.org}

\author{Ridho Reinanda}
\orcid{}
\affiliation{%
\institution{Bloomberg}
\city{London}
\country{U.K.}
}
\email{rreinanda@bloomberg.net}

\author{Abhinav Khaitan}
\orcid{}
\affiliation{%
\institution{Bloomberg}
\city{New York}
\country{U.S.}
}
\email{akhaitan10@bloomberg.net}

\author{Miles Osborne}
\orcid{}
\affiliation{%
\institution{Bloomberg}
\city{London}
\country{U.K.}
}
\email{mosborne29@bloomberg.net}

\author{Giorgio Stefanoni}
\orcid{}
\affiliation{%
\institution{Bloomberg}
\city{London}
\country{U.K.}
}
\email{gstefanoni1@bloomberg.net}

\author{Prabhanjan Kambadur}
\orcid{}
\affiliation{%
\institution{Bloomberg}
\city{New York}
\country{U.S.}
}
\email{pkambadur@bloomberg.net}

\author{Maarten de Rijke}
\orcid{}
\affiliation{%
\institution{University of Amsterdam}
\city{Amsterdam}
\country{The Netherlands}
}
\email{derijke@uva.nl}

\begin{document}

\begin{abstract}
%
%
Knowledge graphs (KGs) model facts about the world;
%
they consist of nodes (entities such as companies and people) that are
connected by edges (relations such as \emph{founderOf}).
Facts encoded in KGs are frequently used by search applications to augment result pages.
When presenting a KG fact to the user, providing other facts that are pertinent to that main fact can enrich the user experience and support exploratory information needs.
%
%
{\em KG fact contextualization} is the task of augmenting a given KG fact with additional and useful KG facts.
%
%
The task is challenging because of the large size of KGs; discovering other relevant facts even in a small neighborhood of the given fact results in an enormous amount of candidates.
%
%
%

%
We introduce a neural fact contextualization method ({\em NFCM}) to address the KG fact contextualization task.
NFCM first generates a set of candidate facts in the neighborhood of a given fact and then ranks the candidate facts using a supervised learning to rank model.
The ranking model combines features that we automatically learn from data and that represent the query-candidate facts with a set of hand-crafted features we devised or adjusted for this task.
In order to obtain the annotations required to train the learning to rank model at scale, we generate
training data automatically using distant supervision on a large entity-tagged text corpus.
We show that ranking functions learned on this data are effective at contextualizing KG facts.
Evaluation using human assessors shows that it significantly outperforms
several competitive baselines.
\end{abstract}

\ccsdesc[300]{Information systems~Presentation of retrieval results}

\keywords{Knowledge graphs, Fact contextualization, Distant supervision}

\maketitle

\section{Introduction}

Knowledge graphs (KGs) have become essential for applications such as search,
query understanding, recommendation and question
answering because they provide a unified view of real-world entities and the facts (i.e., relationships) that hold between them~\cite{blanco2013entity, blanco2015fast, yih2015semantic,
miliaraki2015selena}.
For example,  KGs are increasingly being used to provide direct answers to user queries~\cite{yih2015semantic},
or to construct so-called \emph{entity cards} that provide useful
information about the entity identified in the query.
Recent work~\cite{bota2016playing, hasibi2017dynamic} suggests that search
engine users find entity cards useful and engage with them when they contain
information that is relevant to their search task, for instance in the form of a
set of recommended entities and facts that are related to the
query~\cite{blanco2013entity}.
Previous work has focused on augmenting entity cards with facts that are centered around, i.e., one-hop away from, the main entity of the query~\cite{hasibi2017dynamic}. 
\begin{figure}[t]
	\centering
	 \resizebox{\linewidth}{!}{
      \begin{tikzpicture}[
      align=center,
            > = stealth, 
            shorten > = 1pt, 
            auto,
            node distance = 3cm, 
            semithick 
        ]

        \tikzstyle{every state}=[
            draw = black,
            thick,
            fill = white,
        ]
        \node[state, fill=gray!15] (x) {Bill Gates};
        \node[state, fill=gray!15] (y) [right=2cm of x]{Microsoft};
        \node[state] (s) [below=1.5cm of y]{Software};
         \node[state] (pob) [below left=1.5cm of x] {\small Programmer};
\node[state] (pa) [right=1.5cm of pob] {Paul Allen};
        \node[state] (da) [below right=2cm of y] {1975-04};

    \path[->] (x) edge [font=\bf] node[midway] {\small founderOf} (y);
    \path[->] (y) edge node[midway, sloped, below] {\small industry} (s);
        \path[->] (y) edge node[midway] {\small dateFounded} (da);
        \path[->] (pa) edge node[midway, sloped, below] {\small founderOf} (y);
        \path[->] (x) edge node[midway] {\small profession} (pob);
        \path[->] (pa) edge node[midway] {\small profession} (pob);
    \end{tikzpicture}
    }
     \caption{ A Freebase subgraph that consists of relevant facts to the
         query fact $\mathit{founderOf}(\text{Bill Gates}, \text{Microsoft})$.
    }
	\label{fig:contextualization-graph-gates-microsoft}
\end{figure}
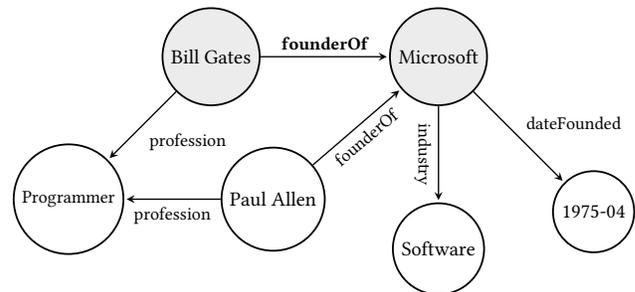
%
%
However, oftentimes a user is interested in KG facts that by definition involve more than one entity (e.g., ``Who founded Microsoft?'' $\longrightarrow$ ``Bill Gates'').
In such cases, we can exploit the richness of the KG by providing query-specific additional facts that increase the user's understanding of the fact as a whole, and that are not necessarily centered around only one of the entities. 
Additional relevant facts for the running example would include Bill Gates' profession, Microsoft's founding date, its main industry and its co-founder Paul Allen (see Figure~\ref{fig:contextualization-graph-gates-microsoft}).
In this case, Bill Gates' personal life 
is less relevant to the fact that he founded Microsoft.

%
Query-specific relevant facts can also be used in other applications to enrich the user experience.
For instance, they can be used to increase the utility of KG question answering (QA) systems that currently only return a single fact as an answer to a natural language question~\cite{yih2015semantic,FNTIR:2016:Bast}.
Beyond QA, systems that focus on automatically generating natural language from KG facts~\cite{EMNLP:2016:Lebret,gardent2017creating} would also benefit from query-specific relevant facts, which can make the generated text more natural and human-like.
This becomes even more important for KG facts that involve tail entities, for which natural language text might not exist for training~\cite{voskarides-generating-2017}.

%
%
%
%
%
%
%
%
In this paper, we address the task of KG fact \emph{contextualization}, that is, given a KG fact that consists of two entities and a relation that connects them, retrieve additional facts from the KG that are relevant to that fact.
%
This task is analogous to ad-hoc retrieval: (i) the ``query'' is a KG fact, (ii) the ``documents'' are other facts in the KG that are in the neighborhood of the ``query''.  
We propose a \emph{neural fact contextualization method} (NFCM), a method that first generates a set of candidate facts that are part of \{1,2\}-hop paths from the entities of the main fact.
NFCM then ranks the candidate facts by how relevant they are for contextualizing the main fact.
We estimate our learning to rank model using supervised data.
The ranking model combines (i) features we automatically learn from data and (ii) those that represent the query-candidate facts with a set of hand-crafted features we devised or adjusted for this task.
Due to the size and heterogeneous nature of KGs, i.e., the large number of entities and relationship types, we turn to distant supervision to gather training data.
Using another, human-verified test collection we gauge the performance of our proposed method and compare it with several baselines.
%
%
%
%
%
We sum up our contributions as follows.
\begin{itemize}
\item We introduce the task of KG fact contextualization where the goal is to, given a fact that consists of two entities and a relationship that connects them, rank other facts from a KG that are relevant to that fact.
\item We propose NFCM, a method to solve KG fact contextualization using distant supervision and learning to rank. Our results show that: (i)~distant supervision is an effective means for gathering training data for this task and
(ii)~a neural learning to rank model that is trained end-to-end outperforms several baselines on a human-curated evaluation set. 
\item We provide a detailed result analysis and insights into the nature of our task.
\end{itemize}
%
%
%
The remainder of the paper is organized as follows.
We first provide a definition of our task in Section~\ref{sec:problem}
and then introduce our method in Section~\ref{sec:method}.
%
%
We describe our experimental setup and detail our results and analyses in Sections~\ref{sec:expsetup} and~\ref{sec:results}, respectively.
We conclude with an overview of related work and an outlook on future directions.
%


\section{Problem statement}
\label{sec:problem}

In this section we provide background definitions and formally define the task of KG fact contextualization.
\subsection{Preliminaries}
\label{sec:preliminaries}
\begin{figure}[t]
	\centering
	\begin{tikzpicture}[
  align=center,
  > = stealth, 
  shorten > = 1pt, 
  auto,
  node distance = 3cm, 
  semithick 
  ]

  \tikzstyle{every state}=[
  draw = black,
  thick,
  fill = white,
  minimum size = 4mm
  ]

  \node[state] (x) {Barack\\Obama};
  \node[state, rectangle] (z) [right of=x] {M1};
  \node[state] (y) [right of=z]{Michelle\\Obama};
  \node[state] (d) [below right=1.6cm of z]{1992-10};
  \node[state] (l) [below right=1.3cm of x]{Hawaii};
  
  \path[->] (x) edge[below]  node[midway] {\small spouse} (z);
  \path[->] (z) edge[below]  node[midway] {\small spouse} (y);
  \path[->] (z) edge[midway]  node[midway, sloped, below] {\small marriageDate} (d);
  \path[->] (x) edge[below]  node[midway, sloped, below] {\small bornIn} (l);
\end{tikzpicture}
	 \caption{
KG subgraph that consists of three facts: $\mathit{bornIn}\langle \text{Barack Obama}, \text{Hawaii}\rangle$, $\mathit{spouseOf}\langle \text{Barack Obama}, \text{Michelle Obama}\rangle$ and $\mathit{marriageDate}\langle \text{M1}, \text{1992-10}\rangle$.
M1 is a CVT entity.
Note that the third fact is an attribute of the second fact.
}
	 \label{rel:spouseOf}
\end{figure}
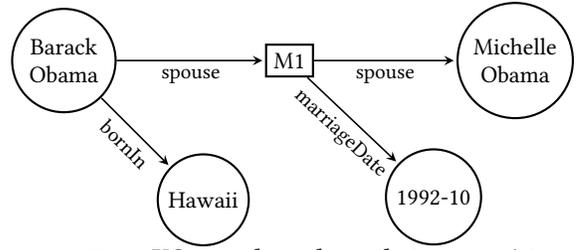

%
%
Let $E = E_n \cup E_c$ be a set of entities, where $E_n$ and $E_c$ are disjoint sets of non-CVT and CVT entities, respectively.\footnote{Compound Value Type (CVT) entities are special entities frequently used in KGs such as Freebase and Wikidata to model fact attributes. See Figure~\ref{rel:spouseOf} for an example.}
Furthermore, let $P$ be a set of predicates.
A \emph{knowledge graph} $K$ is a set of triples $\langle s, p, o \rangle$, where $s, o \in E$ and $p \in P$. By viewing
each triple in $K$ as a labelled directed edge, we can interpret $K$ as a labelled directed graph.
We use Freebase as our knowledge graph~\cite{bollacker2008freebase,nickel2015review}. 
%

%
A path in K is a non-empty sequence $\langle s_0, p_0, t_0 \rangle, \ldots ,\langle s_m, p_m, t_m \rangle$  of triples from K such that $t_i = s_{i+1}$ for each $i \in {0, m-1}$.

We define a \emph{fact} as a path in $K$ that either:
(i)~consists of 1 triple, $s_0 \in E$ and $t_0 \in E_n$ (i.e., $s_0$ may be a CVT entity), or
(ii)~consists of 2 triples, $s_0, t_1 \in E_n$ and $t_0=s_1 \in E_c$ (i.e., $t_0=s_1$ must be a CVT entity).
A fact of type (i) can be an attribute of a fact of type (ii), iff they have a common CVT entity (see Figure~\ref{rel:spouseOf} for an example).

Let $R$ be a set of relationships where a \emph{relationship} $r \in R$ is a label for a set of facts that share the same predicates but differ in at least one entity.
For example, $\mathit{spouseOf}$ is the label of the fact depicted in the top part of Figure~\ref{rel:spouseOf} and consists of two triples.
Our definition of a relationship corresponds to direct relationships between entities, i.e., one-hop paths or two-hop paths through a CVT entity.
For the remainder of this paper, we refer to a specific fact $f$ as $r\langle s, t\rangle$, where $r \in R$ and $s, t \in E$.

\subsection{Task definition}
\label{sec:task_definition}
Given a query fact $f_q$ and a KG $K$, we aim to find a set of other, relevant facts from $K$.
Specifically, we want to enumerate and rank a set of candidate facts $F = \{f_c: f_c \subseteq K, f_c \neq f_q\}$ based on their relevance to $f_q$.
A candidate fact $f_c$ is \emph{relevant} to the query fact $f_q$ if it provides useful and contextual information.
%
Figure \ref{fig:contextualization-graph-gates-microsoft} shows an example part of our KG that is relevant to the query fact $\mathit{founderOf}\langle \text{Bill Gates}, \text{Microsoft} \rangle$.
Note that a candidate fact does not have to be directly connected to both entities of the query fact to be relevant, e.g., $\mathit{profession} \langle \text{Paul Allen}, \text{Programmer}\rangle$.
Similarly, a fact can be related to one or more entities in the relationship instance, e.g., $\mathit{parentOf} \langle \text{Bill Gates}, \text{Jennifer Katharine Gates} \rangle$, but not provide any context, thus being considered irrelevant.







\section{Method}
\label{sec:method}
In this section we describe our proposed neural fact contextualization method (NFCM) which works in two steps.
First, given a query fact $f_q$, we enumerate a set of candidate facts $F = \{f_c: f_c \subseteq K\}$ (see Section~\ref{sec:enumerate-facts}).
Second, we rank the facts in $F$ by relevance to $f_q$ to obtain a final ranked list $F'$ using a supervised learning to rank model (see Section~\ref{sec:ranking-facts}).
We describe how we use distant supervision to automatically gather the required annotations to train the supervised learning to rank model in Section \ref{sec:training-data}.
%

\subsection{Enumerating KG facts}
\label{sec:enumerate-facts}
\begin{algorithm}[t]
\caption{Fact enumeration for a given query fact $f_q$.}
\label{alg:fact-enumeration}
\begin{algorithmic}[1]
\Require{A query fact $f_q=r\langle s, t \rangle$}
\Ensure{A set of candidate facts $F$}
\State $F \gets \{ \} $
\For{$e \in \{s, t\}$}
    \For{$n \in$ \textproc{GetOutNeighbors($e$)} + \textproc{GetInNeighbors($e$)}}
        \State $F.addAll($\textproc{GetFacts($e, n$)}$)$
        \If{\textproc{IsClassOrType($n$)}}
            \State continue
        \EndIf
        \For{$n_2 \in$ \textproc{GetOutNeighbors($n$)}}
            \State $F.addAll($\textproc{GetFacts($n, n_2$)}$)$
        \EndFor
        \For{$n_2 \in$ \textproc{GetInNeighbors($n$)}}
            \State $F.addAll($\textproc{GetFacts($n_2, n$)}$)$
        \EndFor
    \EndFor
\EndFor
\State \Return $F$
\end{algorithmic}
\end{algorithm}
In this section we describe how we obtain the set of candidate facts $F$ from $K$ given a query fact $f_q=r \langle s,t \rangle$.
Because of the large size of real-world KGs---which can easily contain upwards of 50 million entities and 3 billion facts~\cite{pellissier2016freebase}---
it is computationally infeasible to add all possible facts of $K$ in $F$.
Therefore, we limit $F$ to the set of facts that are in the broader neighborhood of the two entities $s$ and $t$.  
Intuitively, facts that are further away from the two entities of the query fact are less likely to be relevant. 

\begin{figure}[t]
	\centering
    \resizebox{\linewidth}{!}{
    	\begin{tikzpicture}[
      align=center,
            > = stealth, 
            shorten > = 1pt, 
            auto,
            node distance = 3cm, 
            semithick 
        ]

        \tikzstyle{every state}=[
            draw = black,
            thick,
            fill = white,
            minimum size = 5mm
        ]

        \node[state, fill=gray!15] (x) {Bill Gates};
                \node[state] (bmgf) [right=1.5cm of x] {B\&M G.\\Foundation};
    \node[state, rectangle, fill=gray!15] (med1)  [below=1cm of bmgf] {$cvt_1$};
    \node[state, fill=gray!15] (y) [right=2cm of bmgf]{Melinda\\Gates};
    \node[state] (m) [below left=2.5cm of x]{Microsoft};
        \node[state] (s) [below left=1.5cm of m]{Software};
        \node[state] (pa) [left=1.5cm of m] {Paul Allen};
        \node[state] (prog) [left=1.5cm of x] {Programmer};
    \node[state] (daf) [below=1.5cm of m] {1975-04};
    \node[state] (dobb) [above left=1.5cm of x] {1955-10};
    \node[state] (dobm) [above=1.5cm of y] {1964-08};
        \node[state] (lanai) [below right=2cm of med1] {Lanai};
        \node[state] (med1from) [right=2cm of med1] {1994-01};

    \node[state] (jkg) [above=1cm of bmgf] {Jennifer\\Gates};

        
        \node[state, rectangle] (med2)  [below=3cm of x] {$cvt_2$};
        \node[state] (chair)  [below left=1cm of med2] {CEO};
        \node[state] (chairfrom)  [right=1cm of med2] {1975-04};
        \node[state] (chairto)  [below right=1cm of med2] {2000-01};
        


        \path[->] (x) edge node[midway, sloped, below] {\small spouse} (med1);
        \path[->] (med1) edge node[midway, sloped, below] {\small spouse} (y);
    \path[->] (x) edge node[midway, sloped, below] {\small founderOf} (m);
    \path[->] (m) edge node[midway, sloped, below] {\small industry} (s);
        
        \path[->] (pa) edge node[midway, sloped, below] {\small founderOf} (m);

        \path[->] (x) edge node[midway] {\small profession} (prog);
        \path[->] (pa) edge node[midway] {\small profession} (prog);
        
        \path[->] (m) edge node[midway, sloped, below] {\small dateFounded} (daf);
        \path[->] (x) edge node[midway, sloped, below] {\small dateOfBirth} (dobb);
        \path[->] (y) edge node[midway, sloped, below] {\small dateOfBirth} (dobm);
        
        \path[->] (med1) edge node[midway, sloped, below] {\small ceremonyAt} (lanai);
        \path[->] (med1) edge node[midway, sloped, below] {\small marriageDate} (med1from);
        
         
         
                  \path[->] (x) edge node[midway, sloped, below] {\small parentOf} (jkg);
         \path[->] (y) edge node[midway, sloped, below] {\small parentOf} (jkg);
         
         \path[->] (x) edge node[midway, sloped, below] {\small founderOf} (bmgf);
         \path[->] (y) edge node[midway, sloped, below] {\small founderOf} (bmgf);

         \path[->] (x) edge node[midway, sloped, below] {\small leaderOf} (med2);
         \path[->] (med2) edge node[midway, sloped, below] {\small leadership} (m);
         \path[->] (med2) edge node[midway, sloped, below] {\small leaders} (chair);         
         \path[->] (med2) edge node[midway, sloped, below] {\small from} (chairfrom);
         \path[->] (med2) edge node[midway, sloped, below] {\small to} (chairto);



    %
    \end{tikzpicture}
        }
     \caption{Graph with a subset of the facts that are enumerated for the query fact $\mathit{spouseOf}(\text{Bill Gates}, \text{Melinda Gates})$. The entities of the query fact are shaded. 
     }
	 \label{fig:candidate-graph-example}
 \end{figure}
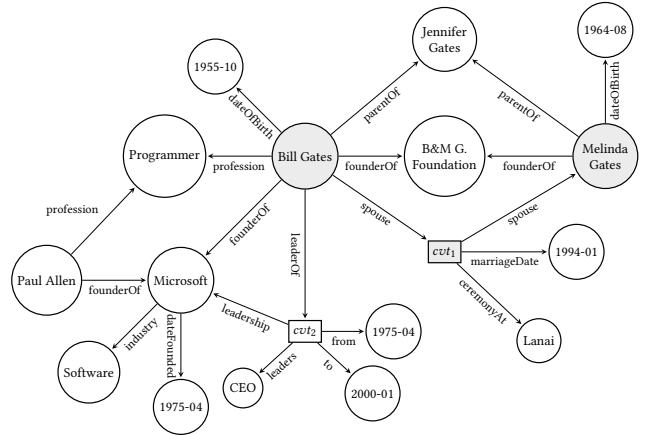

The procedure we follow is outlined in Algorithm~\ref{alg:fact-enumeration}.
This algorithm enumerates the candidate facts for $f_q = r \langle s, t \rangle $ that are at most 2 hops away from either $s$ or $t$.
Three exceptions are made to this rule: (i)~CVT entities are not counted as hops, 
(ii)~we do not include $f_q$ in $F$ as it is trivial,
and (iii)~to reduce the search space, we do not expand intermediate neighbors that
represent an entity class or a type (e.g., ``actor'') as these can have
millions of neighbors.
Figure~\ref{fig:candidate-graph-example} shows an example graph with a subset of the
facts that we enumerate for the query fact $\mathit{spouseOf}\langle \text{Bill
Gates}, \text{Melinda Gates} \rangle$ using Algorithm~\ref{alg:fact-enumeration}.

\subsection{Fact ranking}
\label{sec:ranking-facts}

Next, we describe how we rank the set of enumerated candidate facts $F$ 
with respect to their relevance to the
query fact $f_q=r\langle s, t \rangle$.
The overall methodology is as follows. For each candidate fact $f_c \in F$, we create a
pair $(f_q, f_c)$---an analog to a query-document
pair---and score it using a function $u: (f_q, f_c) \to [0,1] \in R$ (higher values indicate higher relevance).
We then obtain a ranked list of facts $F'$ by sorting the facts in $F$ based on their score.

We begin by describing the training procedure we follow and continue with the network architecture we use for learning our scoring function $u$.

\paragraph{Learning procedure}
\label{sec:learning-procedure}
We train a network that learns the scoring function $u(f_q, f_c)$ end-to-end in mini-batches using stochastic gradient descent (we define the network architecture below).
We optimize the model parameters using Adam~\cite{kingma2014adam}.
During training we minimize a pairwise loss to learn the function $u$, while during inference we use the learned function $u$ to score a query-candidate fact pair ($f_q$, $f_c$).
This paradigm has been shown to outperform pointwise learning methods in ranking tasks, while keeping inference efficient~\cite{dehghani2017neural}.
Each batch $B$ consists of query-candidate fact pairs ($f_q$, $f_c$) of a single query fact $f_q$.
For constructing $B$ for a query fact $f_q$, we use all pairs ($f_q$, $f_c$) that are labeled as relevant and sample $k$ pairs ($f_q$, $f_c$) that are labeled as irrelevant.  
During training, we minimize the mean pairwise squared error between all pairs of ($f_q$, $f_c$) in $B \times B$:
\begin{align}
	L(B, \theta) = \frac{1}{|B|} \sum_{\langle x_1, x_2 \rangle \in B \times B } ([l(x_1) - l(x_2)] - [u(x_1) - u(x_2)] )^2,
\end{align}
where $x_1=(f_q, f_{c_1})$ and $x_2=(f_q, f_{c_2})$ are query-candidate fact pairs in the set $B \times B$,  $l(x) \in \{0, 1\}$ is the relevance label of a query-candidate fact pair $x$, $|B|$ is the batch size, and $\theta$ are the parameters of the model which we define below.

\begin{figure}[t]
	\centering
	\includegraphics[scale=0.45]{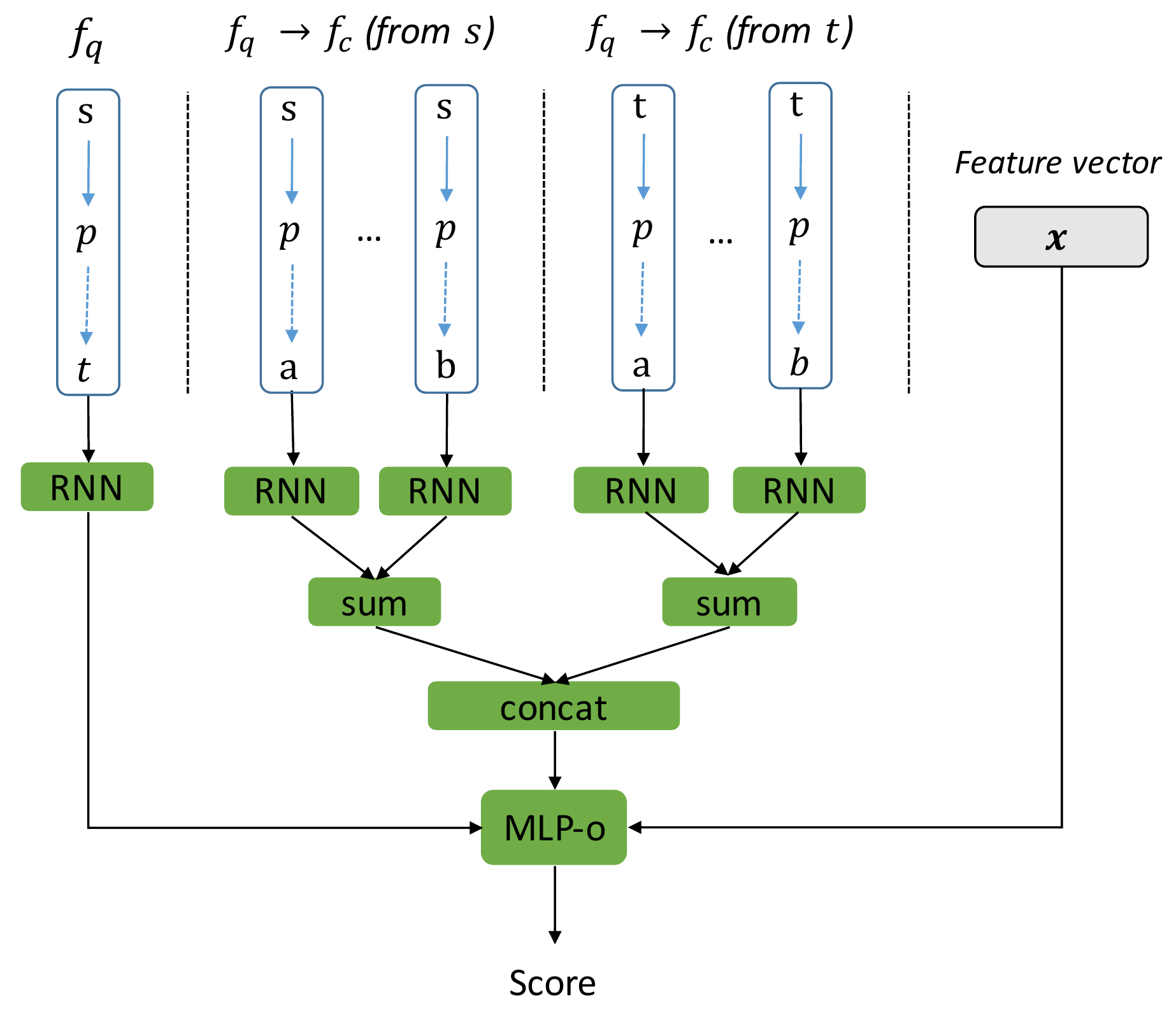}
	 \caption{
Network architecture that learns a scoring function $u(f_q, f_c)$.
Given a query fact $f_q=\mathit{r} \langle \text{s}, \text{t}\rangle$ and a candidate fact $f_c=\mathit{r'} \langle \text{a}, \text{b}\rangle$ it outputs a score $u(f_q, f_c)$.
``$f_q \rightarrow f_c$ (from $e$)'' is a label for the paths that start from an entity $e$ of the query fact (either $s$ or $t$) and end at an entity $e'$ of the candidate fact $f_c$.
Note that $p$ is a variable in this figure, i.e., it might refer to different predicates.
}
	 \label{fig:architecture}
\end{figure}

\paragraph{Network architecture}
Figure~\ref{fig:architecture} shows the network architecture we designed for learning the scoring function $u(f_q, f_c)$.
%
%
%
%
We encode the query fact $f_q$ in a vector $\boldsymbol{v_q}$ using an RNN (see Section~\ref{sec:enc-single-fact}).
%
As we will explain further in that section, we do not model the entities in the facts independently due to the large number of entities; instead, we model each entity as an aggregation of its types.
Therefore, instead of modeling the candidate fact $f_c$ in isolation and losing per-entity information, we first enumerate all the paths up to two hops away from both the entities of the query fact $f_q$ ($s$ and  $t$) to all the entities of the candidate fact $f_c$ ($s'$ and $t'$).
Let $A_s$ denote the set of paths from $s$ to all the entities of $f_c$.
Let $A_t$ denote the set of paths from $t$ to all the entities of $f_c$.
For each $A \in \{A_s, A_t\}$, we first encode all the paths in $A$ using an RNN (Section~\ref{sec:enc-single-fact}), and then combine the resulting encoded paths using the procedure described in Section~\ref{sec:comb-fact}.
We denote the vectors obtained from the above procedure for $A_s$ and $A_t$ as $\boldsymbol{v}_{as}$ and $\boldsymbol{v}_{at}$, respectively.
Then we obtain a vector $\boldsymbol{v}_{a}=[\boldsymbol{v}_{as}, \boldsymbol{v}_{at}]$, where $[\cdot,\cdot]$ denotes the concatenation operation (middle part of Figure~\ref{fig:architecture}).
Note that we use the same RNN parameters for all the above operations.
%
%
To further inform the scoring function, we design a set of hand-crafted features $\boldsymbol{x}$ (right-most part of Figure~\ref{fig:architecture}).
We detail the hand-crafted features in Section~\ref{sec:features}.
%

Finally, $\text{MLP-o}([\boldsymbol{v}_q, \boldsymbol{v}_{a}, \boldsymbol{x}])$ is a multi-layer perceptron with $\alpha$ hidden layers of dimension $\beta$ and one output layer that outputs $u(f_q, f_c)$.
We use a ReLU activation function in the hidden layers and a sigmoid activation function in the output layer.
%
%
%
We vary the number of layers to capture non-linear interactions between the features in $\boldsymbol{v}_q$, $\boldsymbol{v}_a$, and $\boldsymbol{x}$.

The remainder of this section is structured as follows. Section~\ref{sec:enc-single-fact} describes how we encode a single fact, Section~\ref{sec:comb-fact} describes how we combine the representations of a set of facts and finally Section~\ref{sec:features} details the hand-crafted features.

\subsubsection{Encoding a single fact}
\label{sec:enc-single-fact}
Recall from Section~\ref{sec:preliminaries} that a fact $f$ is a path in the KG. 
In order to model paths we turn to neural representation learning.
More specifically, since paths are sequential by nature we employ recurrent neural networks (RNNs) to encode them in a single vector~\cite{guu2015traversing,das2017chains}. 
This type of modeling has proven successful in predicting missing links in KGs~\cite{das2017chains}.
One restriction that we have in modeling such paths is the very large number of entities ($\sim1.5$ million entities in our dataset) and, since learning an embedding for such large numbers of entities requires prohibitively large amounts of memory and data, we represent each entity using an aggregation of its types~\cite{das2017chains}.
Formally, let $\boldsymbol{W}_z$ denote a $|Z| \times d_z$ matrix, where each row is an embedding of an entity type $z$, $|Z|$ is the number of entity types in our dataset and $d_z$ is the entity type embedding dimension.
Let $\boldsymbol{W}_p$ denote a $|P| \times d_p$ matrix, where each row is an embedding of a predicate $p$, $|P|$ is the number of predicates in our dataset, and $d_p$ is the predicate embedding dimension.
In order to model \emph{inverse} predicates in paths (e.g., $\text{Microsoft} \rightarrow \mathit{founderOf}^{-1} \rightarrow \text{Paul Allen}$), we also define a $|P| \times d_p$ matrix $\boldsymbol{W}_{p_i}$, which corresponds to embeddings of the inverse of each predicate~\cite{guu2015traversing}.

The procedure we follow for modeling a fact $f$ is as follows.
For simplicity in the notation, in this Section we denote a path as a sequence of alternate entities and predicates $[s_0, p_0, \ldots t_m]$, instead of a sequence of triples as defined in Section~\ref{sec:preliminaries}.
For each entity $e \in f$, we first retrieve the types of $e$ in $K$.
From these, we only keep the 7 most frequent types in $K$, which we denote as $Z_e$~\cite{das2017chains}.
We then project each $z \in Z_e$ to its corresponding type embedding $\boldsymbol{w}_z \in \boldsymbol{W}_z$ and perform element-wise sum on these embeddings to obtain an embedding $w_e$ for entity $e$. 
We project each predicate $p \in f$ to its corresponding embedding $\boldsymbol{w}_p$ ($\boldsymbol{w}_p \in \boldsymbol{W}_{p_i}$ if $p$ is inverse, $\boldsymbol{w}_p \in \boldsymbol{W}_{p}$ otherwise).
The resulting projected sequence $X_f =[\boldsymbol{w}_{s_0}, \boldsymbol{w}_{p_0} , \ldots, \boldsymbol{w}_{t_m}]$ is passed to a uni-directional recurrent neural network (RNN).
The RNN has a sequence of hidden states $[\boldsymbol{h}_1, \boldsymbol{h}_2, \ldots, \boldsymbol{h}_n]$, where $\boldsymbol{h}_i=tanh(\boldsymbol{W_{hh}} \boldsymbol{h}_{i-1} + \boldsymbol{W_{xh}} \boldsymbol{x}_i)$, and $\boldsymbol{W_{hh}}$ and $\boldsymbol{W_{xh}}$ are the parameters of the RNN.
The RNN is initialized with zero state values.
We use the last state of the RNN $\boldsymbol{h}_n$ as the representation of the fact $f$. 

\subsubsection{Combining a set of facts}
\label{sec:comb-fact}
%
We obtain the representation of the set of encoded facts using element-wise summation of the encoded facts (vectors).
We leave more elaborate methods for combining facts such as attention mechanisms~\cite{bahdanau2014neural,das2017chains} for future work.

\subsubsection{Hand-crafted features}
\label{sec:features}

\begin{table*}[t]
\caption{Notation
}
\label{tab:notation}

\begin{tabularx}{\linewidth}{l p{6cm} p{5cm}}
\toprule
\bf Name & \bf Description & \bf Definition \\
\midrule
$\mathit{NumTriples}$ & Number of triples in $K$ &  $|\{\langle s, p, t\rangle: \langle s, p, t\rangle \in K\} |$  \\
$\mathit{TriplesPred}(p)$ & Set of triples that have predicate $p$ & $\{ \langle s, p', t \rangle :  \langle s, p', t \rangle \in K , p' = p \}$\\

$\mathit{TriplesEnt}(e)$ & Set of triples that have entity $e$ & $\{ \langle s, p, t \rangle :  \langle s, p, t \rangle \in K, s = e \vee t = e \}$ \\

$\mathit{TriplesSubj}(e)$ & Set of triples that have entity $e$ as subject & $\{ \langle s, p, t \rangle :  \langle s, p, t \rangle \in K, s = e\} $\\

$\mathit{TriplesObj}(e)$ & Set of triples that have entity $e$ as object & $\{ \langle s, p, t \rangle :  \langle s, p, t \rangle \in K, t = e\} $\\
$\mathit{UniqEnt}(T)$ & The unique set of entities in a set of triples $T$ & $ \bigcup \{ \{s, t\}: \langle s, p, t \rangle \in T \} $
\\

$\mathit{Types}(e)$ & The set of types of entity $e$ & $\{ z:  \langle e, \mathit{type}, z \rangle \in K\} $ \\



%
$\mathit{Entities}(f)$  & The set of entities of fact $f$ & $  \bigcup \{ \{s, t\}: \forall \langle s, p, t \rangle \in f \}  $ \\
$\mathit{Preds}(f)$  & The set of predicates of fact $f$ & $\{ p:  \langle s, p, t \rangle  \in f \} $\\ 
\bottomrule

\end{tabularx}
\end{table*}

%
%
%
Here, we detail the hand-crafted features $\boldsymbol{x}$ we designed or adjusted for this task.
%
%
Table~\ref{tab:notation} lists the notation we use.
We generate features based on feature templates that are divided into three groups: (i)~those that give us a sense of
\emph{importance} of a fact, (ii)~those that give us a sense of \emph{relevance} of $(f_q, f_c)$,
and (iii)~a set of miscellaneous features.
Note that we use log-computations to avoid underflows.
%

\paragraph{(i) Fact importance}
This group of feature templates give us a sense on how important a fact $f$ is when taking statistics of the knowledge graph $K$ into account at a global level.
Note that we calculate these features for both facts $f_q$ and $f_c$.
The first of these feature templates measures \emph{normalized predicate frequency} of each predicate $p$ that participates in fact $f$ (we also include the minimum, maximum and average value for each fact as metafeatures~\citep{borisov-2016-using}).
This is defined as the ratio of the size of the set of triples that have predicate $p$ in the KG to the total number of triples:
\begin{align}
  \mathit{PredFreq}(p) = \frac{|\mathit{TriplesPred}(p)|}{\mathit{NumTriples}}.
  \label{eq:pred-freq}
\end{align}
The second feature template is the \emph{normalized entity frequency} for each entity $e$ that participates in fact $f$ (we also include the minimum, maximum and average value for each fact as metafeatures).
This is defined as the ratio of the number of triples in which $e$ occurs in the KG over the number of triples in the KG:
\begin{align}
  \mathit{EntFreq}(e) = \frac{|\mathit{TriplesEnt}(e)|}{\mathit{NumTriples}}.
  \label{eq:ent-freq}
\end{align}
The final feature template in this feature group is \emph{path informativeness}, proposed by~\citet{pirro2015explaining}, which we apply for both $f_q$ and $f_c$ (recall from Section~\ref{sec:preliminaries} that a fact $f$ is a path in the KG).
This feature is analog to TF.IDF and aims to estimate the importance of predicates for an entity.
%
%
%
%
%
The informativeness of a path $\pi$ is defined as follows~\cite{pirro2015explaining}:
\begin{align}
  I(\pi) = \frac{1}{2 |\pi|} \sum_{\langle s, p, t\rangle \in \pi} \mathit{PFITF}_{out}(p, s, K) + \mathit{PFITF}_{in}(p, t, K),
  \label{eq:pathInformativeness}
\end{align}
where:
\begin{align}
  \mathit{PFITF}_{x}(p, e, K) = \mathit{PF}_{x}(p, e) * \mathit{ITF}(p), x \in \{in, out\},
    \nonumber
\end{align}
where $\mathit{ITF}(p)$ is the inverse triple frequency of predicate $p$:
\begin{align}
  \mathit{ITF}(p) = \log \frac{\mathit{NumTriples}}{|\mathit{TriplesPred}(p)|},
    \nonumber
\end{align}
$\mathit{PF}_{out}(p, e)$ is the outgoing predicate frequency of $e$ when $p$ is the predicate:
\begin{align}
  \mathit{PF}_{out}(p, e) = \frac{|\mathit{TriplesSubj}(e) \cap \mathit{TriplesPred}(p)|}{|\mathit{TriplesSubj}(e)|},
    \nonumber
\end{align}
and $\mathit{PF}_{in}(p, e)$ is the incoming predicate frequency of $e$ when $p$ is the predicate:
\begin{align}
  \mathit{PF}_{in}(p, e) = \frac{|\mathit{TriplesObj}(e) \cap \mathit{TriplesPred}(p)|}{|\mathit{TriplesObj}(e)|}.
    \nonumber
\end{align}

\paragraph{(ii) Relevance}
This group of feature templates gives us signal on the relevance of a candidate fact $f_c$ w.r.t.\ the query fact $f_q$.
The first of these feature templates measures \emph{entity similarity} for each pair $(e_1, e_2) \in  \mathit{Entities}(f_q) \times \mathit{Entities}(f_c)$ (we also include the minimum, maximum and average entity similarity as metafeatures).
%
%
We measure entity similarity using type-based Jaccard similarity:
\begin{align}
  \mathit{EntTypeSim}(e_1, e_2) = \mathit{JaccardSim}(\mathit{Types}(e_1), \mathit{Types}(e_2)).
  \label{eq:entity-sim-type-jaccard}
\end{align}
%
%
%

%
\noindent
The next feature template in the \emph{relevance} category is \emph{entity distance},
which allows us to reason about the distance of two entities
$(e_1, e_2) \in  \mathit{Entities}(f_q) \times \mathit{Entities}(f_c)$ (we also include the minimum, maximum and average entity distance as metafeatures).
This feature is defined as the length of the shortest path between $e_1$ and $e_2$ in $K$.
%
The intuition is that we can get a signal for the relevance of $f_c$ by measuring how ``close'' the entities in $f_c$ are to the entities of $f_q$ in the KG.

The next set of features measure \emph{predicate similarity} between every pair
of predicates $(p_1, p_2) \in \mathit{Preds}(f_q) \times \mathit{Preds}(f_c)$ (we also include the minimum, maximum and average predicate similarity as metafeatures).
The intuition is that if $f_c$ has predicates that are highly similar to the predicates in $f_q$, then $f_c$ might be relevant to $f_q$.
We measure predicate similarity in two ways.
First, by measuring the co-occurrence of entities that participate in the
predicates $p_1$ and $p_2$:
\begin{align}
  &\mathit{PredCooccSim}(p_1, p_2) = {} 
    \label{eq:pred-sim-ent-cooccur}\\
  &\mathit{JaccardSim}(\mathit{UniqEnt}(\mathit{TriplesPred}(p_1)), \mathit{UniqEnt}(\mathit{TriplesPred}(p_2))).
  \nonumber
\end{align}
For instance, $\mathit{PredCooccSim}(p_1, p_2)$ would be high for $p1 = \mathit{starredIn}$ and $p2 = \mathit{directedBy}$.
%
%
%
Second, by measuring the jaccard similarity of the set of predicates in $f_q$ with the set of predicates in $f_c$~\cite{pirro2015explaining}:
\begin{align}
  &\mathit{SetPredicatesJaccardSim}(f_q, f_c) = {} 
    \label{eq:path-sim-jaccard} \\
  &\mathit{JaccardSim}(\mathit{Preds}(f_q), \mathit{Preds}(f_c)).
  \nonumber
\end{align}
Finally, we add a binary feature 
that captures whether $f_q$ and $f_c$ have the same CVT entity, i.e., $f_c$ is an attribute of $f_q$.

\paragraph{(iii) Miscellaneous}
This set of features includes whether $f_q$ has a CVT entity (same for $f_c$).
We also include whether an entity is a date (for all entities of $f_q$ and $f_c$).
Finally, we include the concatenation of the predicates of $f_q$ as a feature using one-hot encoding.

\section{Experimental setup}
\label{sec:expsetup}

In this section we describe the setup of our experiments that aim to answer the following research questions:

\begin{description}[nosep]

\item[\textbf{RQ1}] How does NFCM perform compared to a set of heuristic baselines on a crowdsourced dataset?

\item[\textbf{RQ2}] How does NFCM perform compared to a scoring function that scores candidate facts w.r.t. a query fact using the relevance labels gathered from distant supervision on a crowdsourced dataset?

\item[\textbf{RQ3}] Does NFCM benefit from both the handcrafted features and the automatically learned features? 

\item[\textbf{RQ4}] What is the per-relationship performance of NFCM?
How does the number of instances per relationship affect the ranking performance? 
\end{description}

\subsection{Knowledge graph}
\label{sec:datasets-kg}
We use the latest edition of Freebase as our knowledge graph~\cite{bollacker2008freebase}.
We include Freebase relations from the following set of domains:
\textit{People, Film, Music, Award, Government, Business, Organization, Education}.
Following previous work~\cite{mintz2009distant}, we exclude triples that have an equivalent reversed triple.


\subsection{Dataset}
\label{sec:dataset}
%
Our dataset consists of query facts, candidate facts, and a relevance label for each query-candidate fact pair.
%
%
%
In order to construct our evaluation dataset we need to start with a set of relationships.
Given that most of our domains are people-centric, we obtain this set by extracting all relationships from Freebase that have an entity of type \emph{Person} as one of the entities.
%
%
In the end, we are left with 65 unique relationships in total (see Table~\ref{tab:relationships} for example relationships).
\begin{table}[t]
\caption{Examples of relationships used in this work.}
\label{tab:relationships}
\begin{tabularx}{\linewidth}{l l}
\toprule
\bf Domain & \bf Relationship
\\
\midrule
People & $\mathit{spouseOf}(\mathit{person}, \mathit{person})$  
\\
& $\mathit{parentOf}(\mathit{person}, \mathit{person})$  
\\
%
%
& $\mathit{educatedAt}(\mathit{person}, \mathit{organization})$  
\\
\midrule
Business & $\mathit{founderOf}(\mathit{person}, \mathit{organization})$  
\\
& $\mathit{boardMemberOf}(\mathit{person}, \mathit{organization})$ 
\\

& $\mathit{leaderOf}(\mathit{person}, \mathit{organization})$ 
\\

%
\midrule
Film & $\mathit{starredIn}(\mathit{person}, \mathit{film})$  
\\
& $\mathit{directorOf}(\mathit{person}, \mathit{film})$ 
\\
%
%
 & $\mathit{producerOf}(\mathit{person}, \mathit{film})$ 
 \\
%
%
%
%
%
%
\bottomrule
\end{tabularx}
\end{table}
We then proceed to gather our set of query facts.
For each relationship, we sample at most 2,000 query facts, provided that they have at least one relevant fact after applying the procedure described in Section~\ref{sec:training-data}.
In total, the dataset contains 62,044 query facts (954.52 on average per relationship).
After gathering query facts for each relation, we enumerate candidate facts for each query fact using the procedure described in Section~\ref{sec:enumerate-facts}.
Finally, we randomly split the dataset per relationship (70\% of the query facts for training, 10\% for validation, 20\% for testing).
%
%
Table~\ref{tab:dataset-stats} shows statistics of the resulting dataset.

\begin{table}[h]
\caption{Statistics of the dataset gathered using distant supervision (see Section~\ref{sec:training-data}).}
\label{tab:dataset-stats}
\begin{tabularx}{\linewidth}{l l l l l l}
\toprule
\bf Part & \bf \# query facts & \multicolumn{4}{c}{\bf \# candidate facts}
\\
& &  average & median & max. & min.
\\
\midrule
Training & 44,632 & 1,420 & 741 & 9,937 & 2 
\\
Validation & \phantom{0}4,983 & 1,424 & 749  & 9,796 & 3
\\
Test & 12,429 &  1,427 & 771 &  9,924 & 3
\\
\bottomrule
\end{tabularx}
\end{table}


Note that we train and tune the fact ranking models with the training and validation sets in Table~\ref{tab:dataset-stats} respectively, using the automatically gathered relevance labels (see Section~\ref{sec:training-data}).
The test set was only used for preliminary experiments (not reported) and for constructing our manually curated evaluation dataset (see Section \ref{sec:evaluation}).
We describe how we automatically gather noisy relevance labels for our dataset in the next section.

\subsection{Gathering noisy relevance labels}
\label{sec:training-data}

%
Gathering relevance labels for our task is challenging due to the size and heterogeneous nature of KGs, i.e., having a large number of facts and relationship types.
Therefore, we turn to distant supervision~\cite{mintz2009distant} to gather relevance labels at scale.
We choose to get a supervision signal from Wikipedia for the following reasons:
(i) it has a high overlap of entities with the KG we use, and 
(ii) facts that are in KGs are usually expressed in Wikipedia articles alongside other, related facts.
%
We filter Wikipedia to select articles whose main entity is in Freebase, and the entity type corresponds to one of the domains listed in Section~\ref{sec:datasets-kg}.
This results in a set of 1,743,191 Wikipedia articles.
The procedure we follow for gathering relevance labels given a query fact $f_q$ and its set of candidate facts $F$ is as follows.
For a query fact $f_q=r \langle s, t\rangle$, we focus on the Wikipedia article of entity $s$.
First, as Wikipedia style guidelines dictate that only the first mention of another entity should be linked, we augment the articles with additional entity links using an entity linking method proposed in~\cite{voskarides-generating-2017}.
Next, we retain only segments of the Wikipedia article that contain references to $t$.
Here, a segment refers to the sentence that has a reference to $t$ and also one sentence before and one after the sentence.
For each such extracted segment, we assume that it expresses the fact $f_q$, which is a common assumption in gathering noisy training data for relation extraction~\cite{mintz2009distant}.
From the segments, we then collect a set of other entities, $O$, that occur in the same sentence that mentions $t$: for computational efficiency, we enforce $|O| \le{} 20$.
Then, we extract facts for all possible pairs of entities $\langle e_1, e_2 \rangle \in \{O \cup \{s, t\}\} \times \{O \cup \{s, t\}\}$.
If there is a single fact $f_c$ in $K$ that connects $e_1$ and $e_2$, we deem $f_c$ relevant for $f_q$.
However, if there are multiple facts connecting $e_1$ and $e_2$ in $K$, the mention of the fact in the specific segment is ambiguous and thus we do not deem any of these facts as relevant~\cite{sorokin2017context}.
The rest of the facts in $F$ are deemed irrelevant for $f_q$.
%
%

%
The distribution of relevant/non-relevant labels in the distantly supervised data is heavily skewed: out of 87,998,956 facts in total, only 225,032 are deemed to be relevant (0.26\%).
%
This is expected since the candidate fact enumeration step can generate thousands of facts for a certain query fact (see Section~\ref{sec:enumerate-facts}).
As a sanity check, we evaluate the performance of our approach to collect distant supervision data by sampling 5 query facts for each relation in our dataset.
For these query facts, we perform manual annotations on the extracted candidate facts that were deemed as relevant by the distant supervision procedure.
We obtain an overall precision of 76\% when comparing the relevance labels of the distant supervision against our manual annotations.
This demonstrates the potential of our distant supervision strategy for creating training data.
%

%

\subsection{Manually curated evaluation dataset}
\label{sec:evaluation}
In order to evaluate the performance of NFCM on the KG fact contextualization task, we perform crowdsourcing to collect a human-curated evaluation dataset.
The procedure we use to construct this evaluation dataset is as follows.
First, for each of the 65 relationships we consider, we sample five query facts of the relationship from the test set (see Section~\ref{sec:dataset}).
Since fact enumeration for a query fact can yield hundreds or thousands of facts (Section~\ref{sec:enumerate-facts}), it is infeasible to consider all the candidate facts for manual annotation.
Therefore, we only include a candidate fact in the set of facts to be annotated if:
(i) the candidate fact was deemed relevant by the automatic data gathering procedure (Section~\ref{sec:training-data}), or
(ii) the candidate fact matches a fact pattern that is built using relevant facts that appear in at least 10\% of the query facts of a certain relationship. An example fact pattern is $\mathit{parentOf} \langle ?, ?\rangle$, which would match the fact $\mathit{parentOf}\langle \text{Bill Gates}, \text{Jennifer Gates} \rangle$. 
%

We use the CrowdFlower platform, and ask the annotators to judge a candidate fact w.r.t.\ its relevance to a query fact.
We provide the annotators with the following scenario (details omitted for brevity):
\begin{quote}
\textit{We are given a specific real-world fact, e.g.,  ``Bill Gates is the founder of Microsoft'', which we call the query fact.
We are interested in writing a description of the query fact (a sentence or a small paragraph).
The purpose of this assessment task is to identify other facts that could be included in a description of the query fact. Note that even though all facts presented for assessment will be accurate, not all will be relevant or equally important to the description of the main fact.}
\end{quote}
We ask the annotators to assess the relevance of a candidate fact in a 3-graded scale:
\begin{itemize}
	\item \emph{very relevant}: I would include the candidate fact in the description of the query fact; the candidate fact provides additional context to the query fact.
	\item \emph{somewhat relevant}: I would include the candidate fact in the description of the query fact, but only if there is space. 
	\item \emph{irrelevant}: I would not include the candidate fact in the description of the query fact.
\end{itemize}
Alongside each query-candidate fact pair, we provide a set of extra facts that could possibly be used to decide on the relevance of a candidate fact.
These include facts that connect the entities in the query fact with the entities in the candidate fact.
For example, if we present the annotators with the query fact 
$\mathit{spouseOf} \langle \text{Bill Gates}$, Melinda Gates$\rangle$
and the candidate fact 
$\mathit{parentOf} \langle \text{Melinda Gates}$, Jennifer Gates$\rangle$
we also show the fact 
$\mathit{parentOf} \langle \text{Bill Gates}$, Jennifer Gates$\rangle$.
%
%

%
Each query-candidate fact pair is annotated by three annotators.
We use majority voting to obtain the gold labels, breaking ties arbitrarily. 
The annotators get a payment of 0.03 dollars per query-candidate fact pair.
By following the crowdsourcing procedure described above, we obtain 28,281 fact judgments for 2,275 query facts (65 relations, 5 query facts each).
Table~\ref{tab:cf-label-distribution} details the distribution of the relevance labels.
One interesting observation is that facts that are attributes of other facts (see Section \ref{sec:preliminaries}) tend to have relatively more relevant judgments than the ones that are not.
This is expected since some of them are attributes of the query fact (e.g., date of marriage for a \textit{spouseOf} query fact).
Finally, Fleiss' kappa is $\kappa$ = 0.4307, which is considered moderate agreement.
Note that all the results reported in Section~\ref{sec:results} are on the manually curated dataset described here.
\begin{table}[t]
\caption{Relevance label distribution of the crowdsourced evaluation dataset.
}
\label{tab:cf-label-distribution}
\begin{tabularx}{\linewidth}{l@{} c c}
\toprule
\bf Relevance & \bf \mbox{}\hspace*{-.35cm}Non-attribute facts (\%) & \bf Attribute facts (\%) \\
\midrule
Irrelevant
&60.86
&34.34
\\
Somewhat relevant
&34.49
&57.81
\\
Very relevant
&\phantom{0}4.63
&\phantom{0}7.84
\\
\bottomrule
\end{tabularx}
\end{table}
  
\paragraph{Evaluation metrics}
We use the following standard retrieval evaluation metrics: MAP, NDCG@5, NDCG@10 and MRR. In the case of MAP and MRR, which expect binary labels, we consider ``very relevant'' and``somewhat relevant'' as ``relevant".
We report on statistical significance with a paired two-tailed t-test.
\subsection{Heuristic baselines}
\label{sec:baselines}
To the best of our knowledge, there is no previously published method that addresses the task introduced in this paper.
Therefore, we devise a set of intuitive baselines that are used to showcase that our task is not trivial.
We derive them by combining features we introduced in Section~\ref{sec:features}.
We define these heuristic functions below:
\begin{itemize}
\item \textit{Fact informativeness (FI).}
Informativeness of the candidate fact $f_c$ \cite[Eq. \ref{eq:pathInformativeness}]{pirro2015explaining}. This baseline is independent of $f_q$.
\item \textit{Average predicate similarity (APS).}
Average predicate similarity of all pairs of predicates $(p_1, p_2) \in \mathit{Preds}(f_q) \times \mathit{Preds}(f_c)$ (Eq. ~\ref{eq:pred-sim-ent-cooccur}). 
The intuition here is that $f_c$ might be relevant to $f_q$ if it contains predicates that are similar to the predicates of $f_q$.
\item \textit{Average entity similarity (AES).}
Average entity similarity of all pairs of entities in $(e_1, e_2) \in \mathit{Entities}(f_q) \times \mathit{Entities}(f_c)$  (Eq. ~\ref{eq:entity-sim-type-jaccard}). 
The assumption here is that $f_c$ might be relevant to $f_q$ if it contains entities that are similar to the entities of $f_q$.
\end{itemize}
%

%

\subsection{Implementation details}
The models described in Section~\ref{sec:ranking-facts} are implemented in TensorFlow v.1.4.1~\cite{abadi2016tensorflow}.
%
Table~\ref{tab:hyperparams} lists the hyperparameters of NFCM.
We tune the variable hyper-parameters of this table on the validation set and optimize for NDCG@5.

\begin{table}[h]
\caption{Hyperparameters of NFCM, tuned on the validation set.}
\label{tab:hyperparams}
\begin{tabularx}{\linewidth}{p{5.5cm} l}
\toprule
\bf Description & \bf Value(s)\\
\midrule
\# negative samples $k$ during training & [1, 10, 100] \\
Learning rate & [0.01, 0.001, 0.0001] \\
$d_z$: entity type embedding dimension & [64, 128, 256] \\
$d_p$: Predicate embedding dimension & [64, 128, 256] \\
RNN cell size & [64, 128, 256] \\
RNN cell dropout & [0.0, 0.2] \\
$\alpha$: \# hidden layers of MLP-o & [0, 1, 2] \\
$\beta$: \# dimension of MLP-o hidden layers& [50, 100] \\
L2 regularization factor for MLP-o kernel & [0.0, 0.1, 0.2] \\
\bottomrule
\end{tabularx}
\end{table}


\section{Results and Discussion}
\label{sec:results}
In this section we discuss and analyze the results of our evaluation, answering the research questions listed in Section~\ref{sec:expsetup}.

%
%
In our first experiment, we compare NFCM to a set of heuristic baselines we derived to answer \textbf{RQ1}.
Table~\ref{tab:results-baselines} shows the results.
We observe that NFCM significantly outperforms the heuristic baselines by a large margin.
We have also experimented with linear combinations of the above heuristics but the performance does not improve over the individual ones and therefore we omit those results.
We conclude that the task we define in this paper is not trivial to solve and simple heuristic functions are not sufficient.
%
%
\begin{table}[t]
\caption{Comparison between NFCM and the heuristic baselines. Significance is tested between NFCM and AES, the best performing baseline. We depict a significant improvement of NFCM over AES for $p<0.05$ as $^\blacktriangle$.}
\label{tab:results-baselines}
\begin{tabularx}{\linewidth}{p{1.8cm} X X X X}
\toprule
\bf Method & MAP & NDCG@5 &  NDCG@10  &  MRR \\
\midrule
FI
&0.1222
&0.0978
&0.1149
&0.1928
\\
APS
&0.2147
&0.2175
&0.2354
&0.3760
\\
AES
& 0.2950
& 0.3284
& 0.3391
& 0.5214
\\
\midrule
NFCM
& \bf 0.4874$^\blacktriangle$
& \bf 0.5110$^\blacktriangle$
& \bf 0.5289$^\blacktriangle$
& \bf 0.7749$^\blacktriangle$
\\
\bottomrule
\end{tabularx}
\end{table}

%
In our second experiment we compare NFCM with distant supervision and aim to answer \textbf{RQ2}. That is, how does NFCM perform compared to DistSup, a scoring function that scores candidate facts w.r.t. a query fact using the relevance labels gathered from distant supervision.
The aim of this experiment is to investigate whether it is beneficial to learn ranking functions based on the signal gathered from distant supervision, and to see if we can improve performance over the latter.
Table~\ref{tab:results-distsup} shows the results.
We observe that NFCM significantly outperforms DistSup on MAP, NDCG@5, and NDCG@10 and conclude that learning ranking functions (and in particular NFCM) based on the signal gathered from distant supervision is beneficial for this task.
We also observe that NFCM performs significantly worse than DistSup on MRR.
One possible reason for this is that NFCM returns facts that are indeed relevant but were not selected for annotation and thus assumed not relevant, since the data annotation procedure is biased towards DistSup (see Section ~\ref{sec:evaluation}).
We aim to validate this hypothesis by conducting an additional user study in future work.
Nevertheless, having an automatic method for KG fact contextualization trained with distant supervision becomes increasingly important for tail entities for which we might only have information in the KG itself and not in external text corpora or other sources.
%
%
\begin{table}[t]
\caption{Comparison between NFCM and the distant supervision baseline. We depict a significant improvement of NFCM over DistSup as $^\blacktriangle$ and a significant decrease as $^\blacktriangledown$ ($p<0.05$).}
\label{tab:results-distsup}
\begin{tabularx}{\linewidth}{p{1.8cm} X X X X X}
\toprule
\bf Method & MAP & NDCG@5 &  NDCG@10  &  MRR \\
\midrule
DistSup
&0.2831
&0.4489
&0.3983
&\bf 0.8256
\\
NFCM
& \bf 0.4874$^\blacktriangle$
& \bf 0.5110$^\blacktriangle$
& \bf 0.5289$^\blacktriangle$
& 0.7749$^\blacktriangledown$
\\
\bottomrule
\end{tabularx}
\end{table}

\begin{table}[t]
\caption{Comparison between the full NFCM model and its variations. Significance is tested between NFCM and its best variation (LF). We depict a significant improvement of NFCM over LF for $p<0.05$ as $^\blacktriangle$.}
\label{tab:results-constituent}
\begin{tabularx}{\linewidth}{p{1.8cm} X X X X}
\toprule
\bf Method & MAP & NDCG@5 &  NDCG@10 &  MRR \\
\midrule
HF
&0.4620
&0.4753
&0.4989
&0.7180
\\
LF
& 0.4676
& 0.4993
& 0.5134
& 0.7647
\\
\midrule
NFCM
& \bf 0.4874$^\blacktriangle$
& \bf 0.5110
& \bf 0.5289$^\blacktriangle$
& \bf 0.7749
\\
\bottomrule
\end{tabularx}
\end{table}

%
\begin{figure}[t]
	\centering
	\includegraphics[width=\linewidth]{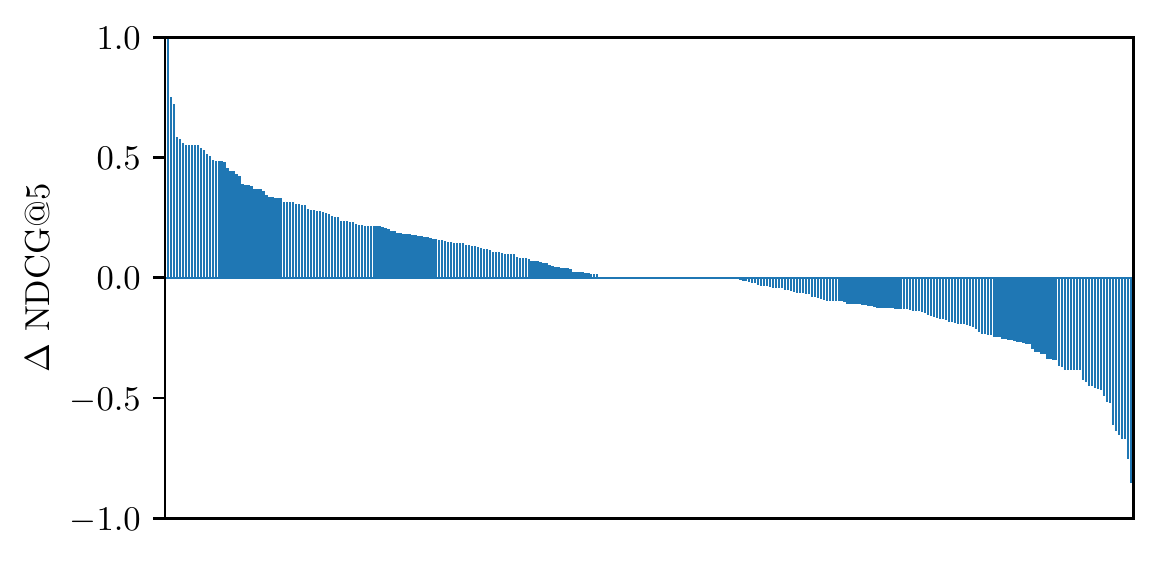}
	 \caption{
	 Per query fact differences in NDCG@5 between the variation of NFCM that only uses the learned features (LF) and the best-performing variation of NFCM that only uses the hand-crafted features (HF). A positive value indicates that LF performs better than HF on a query fact and vice versa.
	 }
	 \label{fig:LF_minus_HF}
\end{figure}

In order to answer \textbf{RQ3}, that is, whether NFCM benefits from both the hand-crafted features and the learned features, we perform an ablation study.
Specifically, we test the following variations of NFCM that only modify the final layer of the architecture (see Section~\ref{sec:ranking-facts}):
\begin{enumerate}[label=(\roman*)]
\item LF: Keeps the learned features ($\boldsymbol{v_q}$ and $\boldsymbol{v_{a}}$), and ignores the hand-crafted features $\boldsymbol{x}$.
\item HF: Keeps the hand-crafted features ($\boldsymbol{x}$) and ignores the learned features ($\boldsymbol{v_q}$ and $\boldsymbol{v_{a}}$).
\end{enumerate}
We tune the parameters of LF and HF on the validation set.
Table~\ref{tab:results-constituent} shows the results.
First, we observe that NFCM outperforms HF by a large margin. Also, NFCM outperforms LF on all metrics (significantly so for MAP and NDCG@10) which means that by combining HF and LF we are able to obtain more relevant results at lower positions of the ranking.  
We aim to explore more sophisticated ways of combining LF and HF in future work.  
In order to verify whether LF and HF have complementary signals, we plot the per-query differences in NDCG@5 for LF and HF in Figure~\ref{fig:LF_minus_HF}.
We observe that the performance of LF and HF varies across query facts, confirming the hypotheses that LF and HF yield complementary signals.
%

%

%
\begin{figure}[t]
	\centering
	\includegraphics[width=\linewidth]{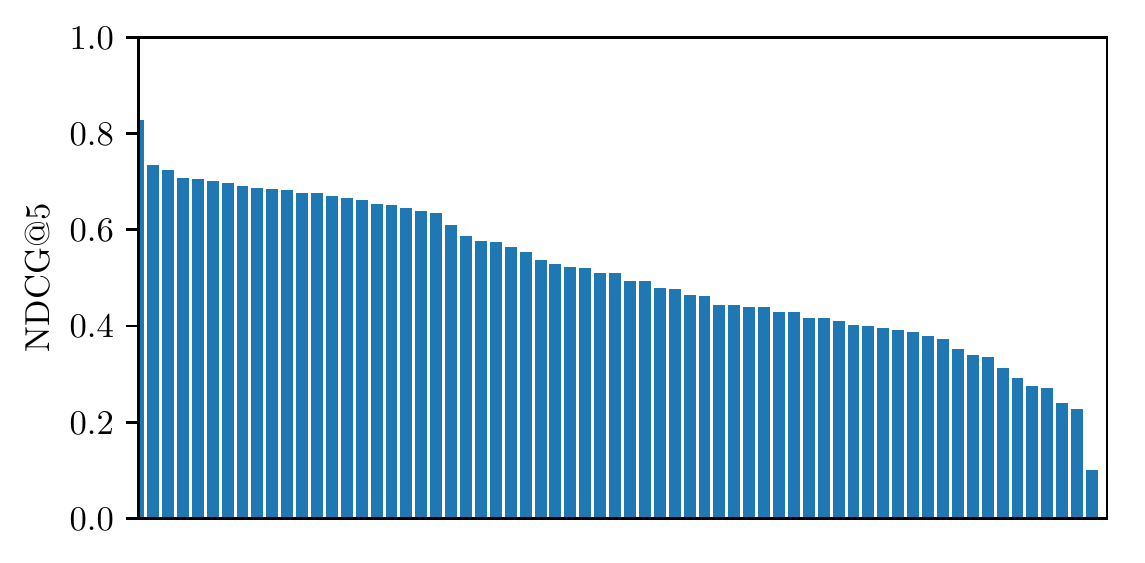}
	 \caption{
	 NDCG@5 for NFCM per relationship.
	 }
	 \label{fig:NFCM_per_relationship}
\end{figure}

In order to answer \textbf{RQ4}, we conduct a performance analysis per relationship.
Figure~\ref{fig:NFCM_per_relationship} shows the per-relationship NDCG@5 performance of NFCM -- query fact scores are averaged per relationship.
The relationship for which NFCM performs best is $\mathit{profession}$, which has a NDCG@5 score of 0.8275.
The relationship for which NFCM performs worst at is $\mathit{awardNominated}$, which has a NDCG@5 score of 0.1.
Further analysis showed that $\mathit{awardNominated}$ has a very large number of candidate facts on average, which might explain the poor performance on that relationship.
%
%

%
\begin{figure}[t]
	\centering
	\includegraphics[width=\linewidth]{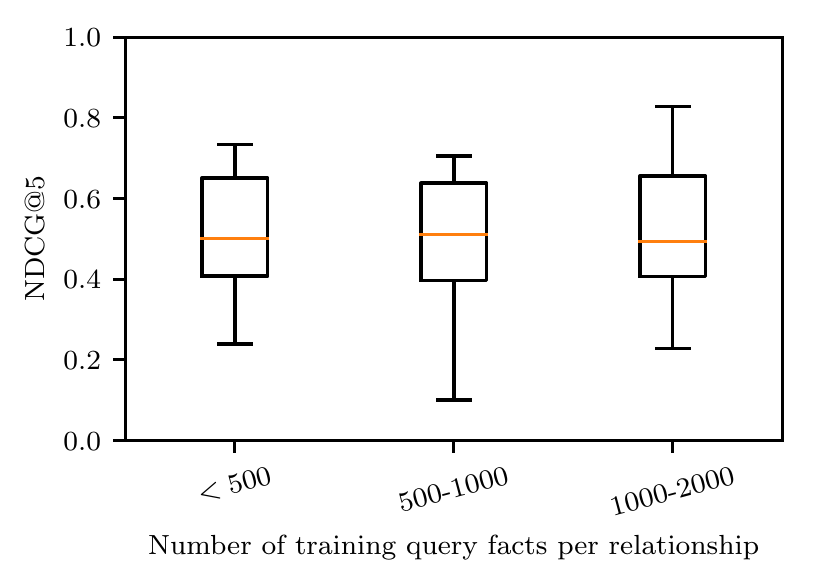}
	 \caption{
	 Box plot that shows NDCG@5 per number of training query facts of each relationship (binned).
	 Each box shows the median score with an orange line and the upper and lower quartiles (maximum and lower values shown outside each box).
	 }
	 \label{fig:per_relationship_numqueries}
\end{figure}
Furthermore, we investigate how the number of queries we have in the training set for each relationship affects the ranking performance.
Figure~\ref{fig:per_relationship_numqueries} shows the results.
From this figure we conclude that there is no clear relationship and thus that NFCM is robust to the size of the training data for each relationship.

%
\begin{figure}[t]
	\centering
	\includegraphics[width=\linewidth]{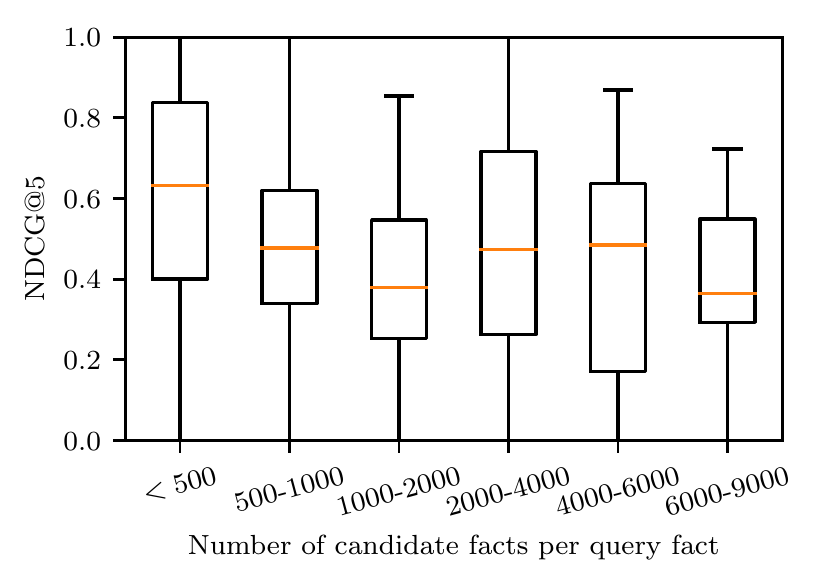}
	 \caption{
	 Box plot that shows NDCG@5 per number of candidate facts of each query fact (binned).
	 Each box shows the median score with an orange line and the upper and lower quartiles (maximum and lower values shown outside each box).
	 }
	 \label{fig:ndcg_5_pernumcandidates}
\end{figure}
Next, we analyse the performance of NFCM with respect to the number of candidates per query fact; Figure~\ref{fig:ndcg_5_pernumcandidates} shows the results.
We observe that the performance decreases when we have more candidate facts for a query, although not by a large margin, and that there does not seem to be a clear relationship between performance and the number of candidates to rank.

\section{Related work}

The specific task we introduce in this paper has not been addressed before, but there is related work in three main areas: entity relationship explanation, distant supervision, and fact ranking.

\subsection{Relationship Explanation}

Explanations for relationships between pairs of entities can be provided in two ways: \emph{structurally}, i.e., by providing paths or sub-graphs in a KG containing the entities, or \emph{textually}, by ranking or generating text snippets that explain the connection.

\citet{fang2011rex} focus on explaining connections between entities by mining relationship explanation patterns from the KG. Their approach consists of two main components: explanation enumeration and explanation ranking. The first phase generates all patterns in the form of paths connecting the two entities in the KG, which are then combined to form explanations. In the final stage, the candidate explanations are ranked using notions of interestingness.
%
\citet{seufert2016espresso} propose a similar approach for entity sets. Their method focuses on explaining the connections between entity sets based on the concept of relatedness cores, i.e., dense subgraphs that have strong relations with both query sets. 
%
\citet{pirro2015explaining} also provide explanations of the relation between entities in terms of the top-k most informative paths between a query pair of entities; such paths are ranked and selected based on path informativeness and diversity, and pattern informativeness. 

As to textual explanations for entity relationships, \citet{voskarides-learning-2015} focus on human-readable descriptions. They model the task as a learning to rank problem for sentences and employ a rich set of features.
\citet{huang2017learning} build on the aforementioned work and propose a pairwise ranking model that leverages clickthrough data and uses a convolutional neural network architecture.
%
%
%
%
%
While these approaches rank existing candidate explanations, \citet{voskarides-generating-2017} focus on generating explanations from scratch. They automatically identify the most common sentence templates for a particular relationship and, for each new relationship instance, these templates are ranked and instantiated using contextual information from the KG. 
%

The work described above focuses on explaining entity relationships in KGs; no previous work has focused on ranking additional KG facts for an input entity relationship as we do in this paper.

\subsection{Distant Supervision}

When obtaining labeled data is expensive, training data can be generated automatically.
%
\citet{mintz2009distant} introduce distant supervision for relation extraction; for a pair of entities that is connected by a KG relation, they treat all sentences that contain those entities in a text corpus as positive examples for that relation.
Follow-up work on relation extraction address the issue of noise related to distant supervision. \citet{riedel-2010-modeling, Surdeanu2012, alfonseca-2012-pattern} refine the model by relaxing the assumptions in the original method or by modeling noisy labels.

Beyond relation extraction, distant supervision has also been applied in other KG-related tasks.
\citet{ren-2015-clustype} introduce a joint approach entity recognition and classification based on distant supervision.
\citet{ling2012fine} used distant supervision to automatically label data for fine-grained entity recognition.

\subsection{Fact Ranking}

In fact ranking, the goal is to rank a set of attributes with respect to an entity. \citet{hasibi2017dynamic} consider fact ranking as a component for entity summarization for entity cards. They approach fact ranking as a learning to rank problem. They learn a ranking model based on importance, relevance, and other features relating a query and the facts. \citet{meza2005ranking} explore a similar task, but rank facts with respect to a pair of entities to discover paths that contain informative facts between the pair.

Graph matching involves matching two graphs and discovering the patterns of relationships between them to infer their similarity \citep{cho2013}. Although our task can be considered as comparing a small query subgraph (i.e., query triples) and a knowledge graph, the goal is different from graph matching which mainly concerns aligning two graphs rather than enhancing one query graph.

\medskip\noindent%
Our work differs from the work discussed above in the following major ways. First, we enrich a query fact between two entities by providing relevant additional facts in the context of the query fact, taking into account both the entities and the relation of the query fact. Second, we rank whole facts from the KG instead of just entities. Last, we provide a distant supervision framework for generating the training data so as to make our approach scalable. 

\section{Conclusion}
In this paper, we introduced the knowledge graph fact contextualization task and proposed NFCM, a weakly-supervised method to address it.
NFCM first generates a candidate set for a query fact by looking at 1 or 2-hop neighbors and then ranks the candidate facts using supervised machine learning. 
NFCM combines handcrafted features with features that are automatically identified using deep learning.
We use distant supervision to boost the gathering of training data by using a large entity-tagged text corpus that has a high overlap with entities in the KG we use.
Our experimental results show that (i)~distant supervision is an effective means for gathering training data for this task, (ii)~NFCM significantly outperforms several heuristic baselines for this task, and (iii)~both the handcrafted and automatically-learned features contribute to the retrieval effectiveness of NFCM.
For future work, we aim to explore more sophisticated ways of combining handcrafted with automatically learned features for ranking.
Additionally, we want to explore other data sources for gathering training data, such as news articles and click logs.
Finally, we want to explore methods for combining and presenting the ranked facts in search engine result pages in a diversified fashion.

\subsection*{Data}
To facilitate reproducibility of our results, we share the 
data used to run our experiments at \href{https://www.techatbloomberg.com/research-weakly-supervised-contextualization-knowledge-graph-facts/}{https://www.techatbloomberg.com/research-weakly-supervised-contextualization-knowledge-graph-facts/}.

\subsection*{Acknowledgements}
The authors would like to thank the anonymous reviewers (and especially reviewer \#1) for their useful and constructive feedback.
This research was supported by
Ahold Delhaize,
Amsterdam Data Science,
the Bloomberg Research Grant program,
the China Scholarship Council,
the Criteo Faculty Research Award program,
Elsevier,
the European Community's Seventh Framework Programme (FP7/2007-2013) under
grant agreement nr 312827 (VOX-Pol),
the Google Faculty Research Awards program,
the Microsoft Research Ph.D.\ program,
the Netherlands Institute for Sound and Vision,
the Netherlands Organisation for Scientific Research (NWO)
under pro\-ject nrs
CI-14-25, 
652.\-002.\-001, 
612.\-001.\-551, 
652.\-001.\-003, 
and
Yandex.
All content represents the opinion of the authors, which is not necessarily shared or endorsed by their respective employers and/or sponsors.

\bibliographystyle{ACM-Reference-Format.bst}
\bibliography{bibliography}

\end{document}